\documentclass{PoS}
\rightline{\parbox{4cm}{\large\rm
DESY 10-203\\
Edinburgh 2010/31\\
LTH 890
}}

\usepackage{amsmath}
\usepackage{amssymb}
\usepackage{graphicx}
\usepackage{wrapfig}
\newcommand{\be}{\begin{equation}}
\newcommand{\ee}{\end{equation}}
\newcommand{\bc}{\begin{center}}
\newcommand{\ec}{\end{center}}

\title{Hyperon Form Factors from $N_f=2+1$ QCD}

\ShortTitle{Hyperon Form Factors}

\author{M.~G\"ockeler$^{a}$, Ph.~H\"agler$^{a}$, R.~Horsley$^{b}$, Y.~Nakamura$^{a,c}$,
  D.~Pleiter$^{d}$, P.~E.~L.~Rakow$^{e}$, A.~Sch\"afer$^{a}$, G.~Schierholz$^{d}$,
  H.~St\"uben$^{f}$, F.~Winter$^{a}$, \speaker{J.~M.~Zanotti}$^{,b}$\\
        \llap{$^a$} Institut f\"ur Theoretische Physik,
                    Universit\"at Regensburg,
                    93040 Regensburg, Germany \\
        \llap{$^b$} School of Physics and Astronomy,
                    University of Edinburgh,
                    Edinburgh EH9 3JZ, UK \\
        \llap{$^c$} Center for Computational Sciences, 
                    University of Tsukuba,
                    Tsukuba, Ibaraki 305-8577, Japan\footnote{present
                      address} \\
        \llap{$^d$} Deutsches Elektronen-Synchrotron DESY,
                    15738 Zeuthen, Germany \\
        \llap{$^e$} Theoretical Physics Division,
                    Department of Mathematical Sciences,
                    University of Liverpool,
                    Liverpool L69 3BX, UK \\
        \llap{$^f$} Konrad-Zuse-Zentrum f\"ur Informationstechnik Berlin,
                    14195 Berlin, Germany \\
        E-mail: \email{jzanotti@ph.ed.ac.uk}}

\author{QCDSF/UKQCD Collaboration}

\abstract{We present results from the QCDSF/UKQCD collaboration for
  the electromagnetic and semi-leptonic form factors for the hyperons.
  The simulations are performed on our new ensembles generated with
  2+1 flavours of dynamical ${\cal O}(a)$-improved Wilson fermions. 
  A unique feature of these configurations is that the quark masses
  are tuned so that the singlet quark mass is held fixed at its
  physical value. 
  We use 5 such choices of the individual quark masses on $24^3\times
  48$ lattices with a lattice spacing of about 0.078~fm.}

\FullConference{The XXVIII International Symposium on Lattice Field Theory, Lattice2010\\
		June 14-19, 2010\\
		Villasimius, Italy}

\begin{document}

\section{Introduction}

The study of the electromagnetic (EM) properties of hadrons provides
important insights into the non-perturbative structure of QCD.
The EM form factors reveal information on the internal structure of
hadrons including their size, charge distribution and magnetisation.

While the EM form factors of the nucleon have received a lot of recent
attention in lattice simulations (see, e.g., \cite{Hagler:2009ni} for
a review), the investigation of the hyperon EM form factors has
so far received only limited attention
\cite{Boinepalli:2006xd,Lin:2008mr}.
These, however, are of significant interest as they provide valuable
insights into the environmental sensitivity of the distribution of
quarks inside a hadron.
For example, how does the distribution of $u$ quarks in $\Sigma^+$
change as we change the mass of the (spectator) $s$ quark?

Semileptonic form factors of the hyperons provide an alternative
method to the standard $K_{\ell 3}$ decays (see
e.g. \cite{Cabibbo:2003cu}) for determining the CKM matrix element,
$|V_{us}|$.
This is done by using the experimental value for the decay rate of the
hyperon beta decays, $B\to b\ell\nu$
\begin{equation}
  \label{eq:vus}
  \Gamma = \frac{G^2_F}{60\pi^3} 
  (M_B-M_b)^5 (1-3\delta)
  |V_{us}|^2|f_1(0)|^2\bigg( 1+3\left|\frac{g_1(0)}{f_1(0)}\right|^2 +
  \cdots \bigg)\ ,
\end{equation}
where $G_F$ is the Fermi constant, $\delta=(M_B-M_b)/(M_B+M_b)$
describes the size of $SU(3)_{\mathrm{flavour}}$ breaking and the
ellipsis denotes terms which are ${\cal O}(\delta^2)$ and can be
safely ignored \cite{Gaillard:1984ny}.
Hence for a determination of $|V_{us}|$, we need to know the form
factors, $f_1(q^2)$ and $g_1(q^2)$, at zero momentum transfer
$(q^2=0)$.
These can be determined on the lattice and are the subject of the
second part of this talk.
Earlier quenched and $N_f=2$ results for $\Sigma^-\to n\ell\nu$ and
$\Xi^0\to\Sigma^+\ell\nu$ can be found in
\cite{Guadagnoli:2006gj,Sasaki:2008ha}.

In this talk we present preliminary results from the QCDSF/UKQCD
Collaboration for the octet hyperon electromagnetic and semi-leptonic
decay form factors determined from $N_f=2+1$ lattice QCD.

%
\section{Simulation Details}
\label{sec:simul}
%
Our gauge field configurations have been generated with $N_f=2+1$
flavours of dynamical fermions, using the tree-level Symanzik improved
gluon action and nonperturbatively ${\cal O}(a)$ improved Wilson
fermions \cite{Cundy:2009yy}.
We choose our quark masses by first finding the
$SU(3)_{\mathrm{flavour}}$-symmetric point where flavour singlet
quantities take on their physical values and vary the individual quark
masses while keeping the singlet quark mass
$\overline{m}_q=(m_u+m_d+m_s)/3=(2m_l+m_s)/3$ constant
\cite{Bietenholz:2010jr}.
Simulations are performed on lattice volumes of $24^3\times 48$ with
lattice spacing, $a=0.078(3)$.
%
%
A summary of the parameter space spanned by our dynamical
configurations can be found in Table~\ref{tab:results}.
More details regarding the tuning of our simulation parameters are
given in Ref.~\cite{Bietenholz:2010jr}.
\begin{table*}
\begin{tabular}{l|c|c|c|c|c|c|c}
Ensemble & $\kappa_l$ & $\kappa_s$ & $m_\pi$\,[MeV]& $m_K$\,[MeV] &
$m_N$\,[GeV] & $m_\Sigma$\,[GeV] & $m_\Xi$\,[GeV]\\
\hline
1 & 0.12083 & 0.12104 & 481 & 420 & 1.257 & 1.209 & 1.180 \\
2 & 0.12090 & 0.12090 & 443 & 443 & 1.231 & 1.231 & 1.231 \\
3 & 0.12095 & 0.12080 & 414 & 459 & 1.205 & 1.240 & 1.258 \\
4 & 0.12100 & 0.12070 & 377 & 473 & 1.175 & 1.242 & 1.278 \\
5 & 0.12104 & 0.12062 & 350 & 485 & 1.123 & 1.222 & 1.280
\end{tabular}
\caption{Pion, Kaon and octet baryon masses on $24^3\times 48$
  lattices with lattice spacing, $a=0.078(3)$\,fm}
\label{tab:results}
\end{table*}

\section{Electromagnetic Form Factors}

On the lattice, we determine the form factors $F_1(q^2)$ and
$F_2(q^2)$ by calculating the following matrix element of the
electromagnetic current
\be
\langle B(p',\,s')| j_{\mu}(q)|B(p,\,s)\rangle
\, = 
 \bar{u}(p',\,s')
 \left[ \gamma_\mu F_1(q^2) +
       \sigma_{\mu\nu}\frac{q_\nu}{2M_B}F_2(q^2) \right] 
 u(p,\,s) \, ,
\label{eq:em-me}
\ee
where $u(p,\,s)$ is a Dirac spinor with momentum, $p$, and spin
polarisation, $s$, $q = p' - p$ is the momentum transfer, $M_B$ is the
mass of the baryon, $B$, and $j_\mu$ is the electromagnetic current.
The Dirac $(F_1)$ and Pauli $(F_2)$ form factors of the proton are
obtained by using $j_\mu^{(p)} = \frac{2}{3}\bar{u}\gamma_\mu u -
\frac{1}{3}\bar{d}\gamma_\mu d$, while the form factors for the
$\Sigma$ and $\Xi$ baryons are obtained through the appropriate
substitution, $u\to s$ or $d\to s$.
It is common to rewrite the form factors $F_1$ and $F_2$ in terms of
the electric and magnetic Sachs form factors, 
$G_e= F_1 + q^2/(2M_N)^2\, F_2$ and $G_m= F_1 + F_2$.

If one is using a conserved current, then (e.g. for the proton)
$F_1^{(p)}(0) = G_e^{(p)}(0) =1$ gives the electric charge,
while $G_m^{(p)}(0) = \mu^{(p)} = 1 + \kappa^{(p)}$
gives the magnetic moment, where $F_2^{(p)}(0) = \kappa^{(p)}$ is the
anomalous magnetic moment.
From Eq.~(\ref{eq:em-me}) we see that $F_2$ always appears with a
factor of $q$, so it is not possible to extract a value for $F_2$ at
$q^2=0$ directly from our lattice simulations.
Hence we are required to extrapolate the results we obtain at finite
$q^2$ to $q^2=0$.
Form factor radii, $r_i=\sqrt{\langle r_i^2\rangle}$, are defined from
the slope of the form factor at $q^2=0$.

In this talk, we are primarily interested in searching for any
$SU(3)$-flavour breaking effects in the octet hyperon form factors.
In order to highlight these effects, we will consider ratios of the
individual quark contributions to the hyperon radii.
For example, the ratio $\langle r^2_1\rangle_{u_\Sigma}/\langle
r^2_1\rangle_{u_p}$ will tell us about the distribution of the doubly
represented ($u$-)quark in a baryon as we change the doubly-singly
represented quark mass splitting (in effect, changing the mass of the
spectator quark).

\begin{figure}[t]
     \vspace*{-2mm}
     \hspace{-0.2mm}
   \begin{minipage}{0.47\textwidth}
     \hspace{-6mm}
          \includegraphics[clip=true,width=1.12\textwidth]{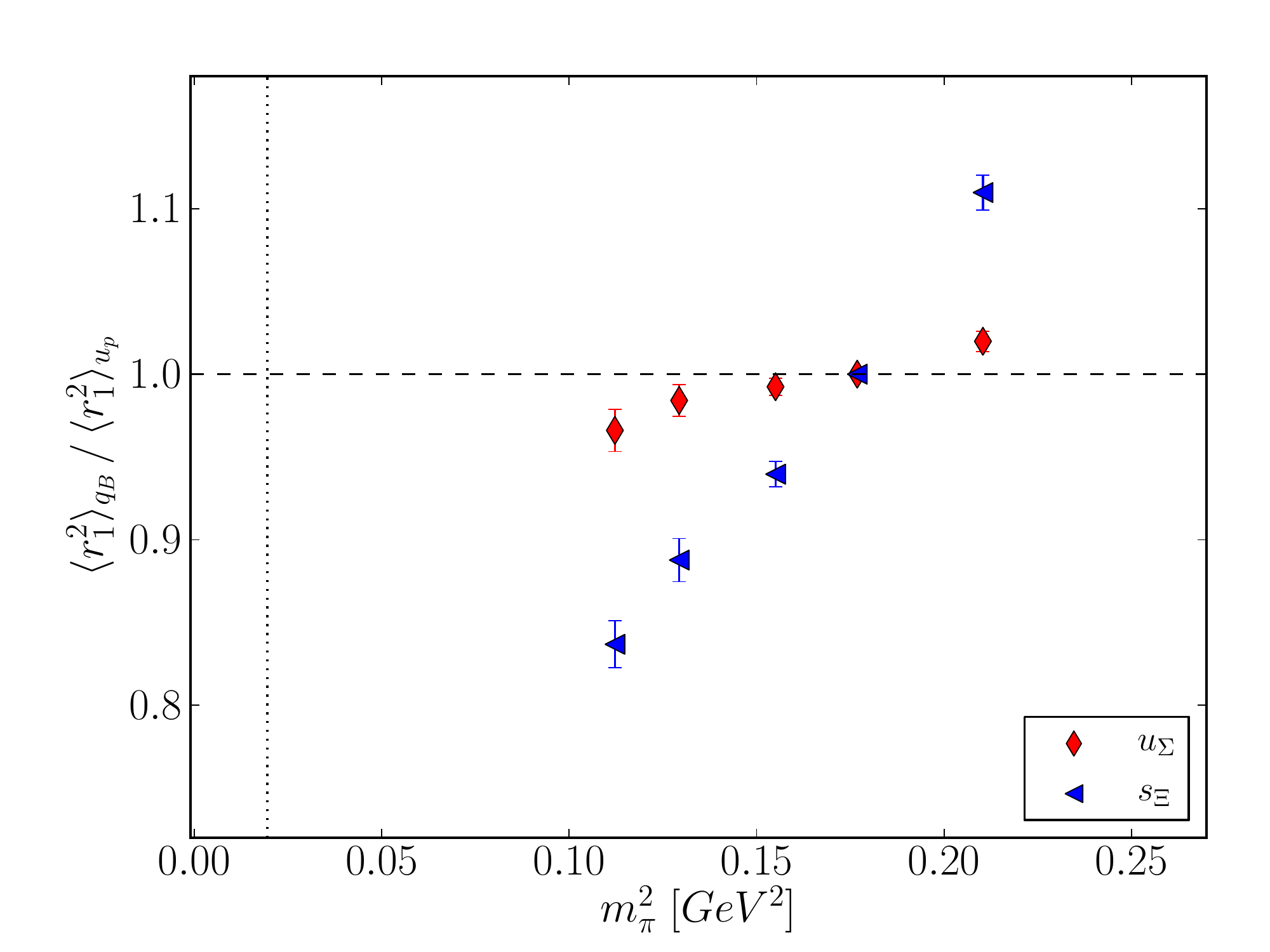}
          \caption{Results for the ratio of the doubly-represented
            quark's contribution to the Dirac radius of the hyperons,
            $\langle r^2_1\rangle_{q_B}/\langle
            r^2_1\rangle_{u_p}$.}
\label{fig:r1D}
     \end{minipage}
     \hspace{3mm}
    \begin{minipage}{0.47\textwidth}
     \hspace{-5mm}
          \includegraphics[clip=true,width=1.12\textwidth]{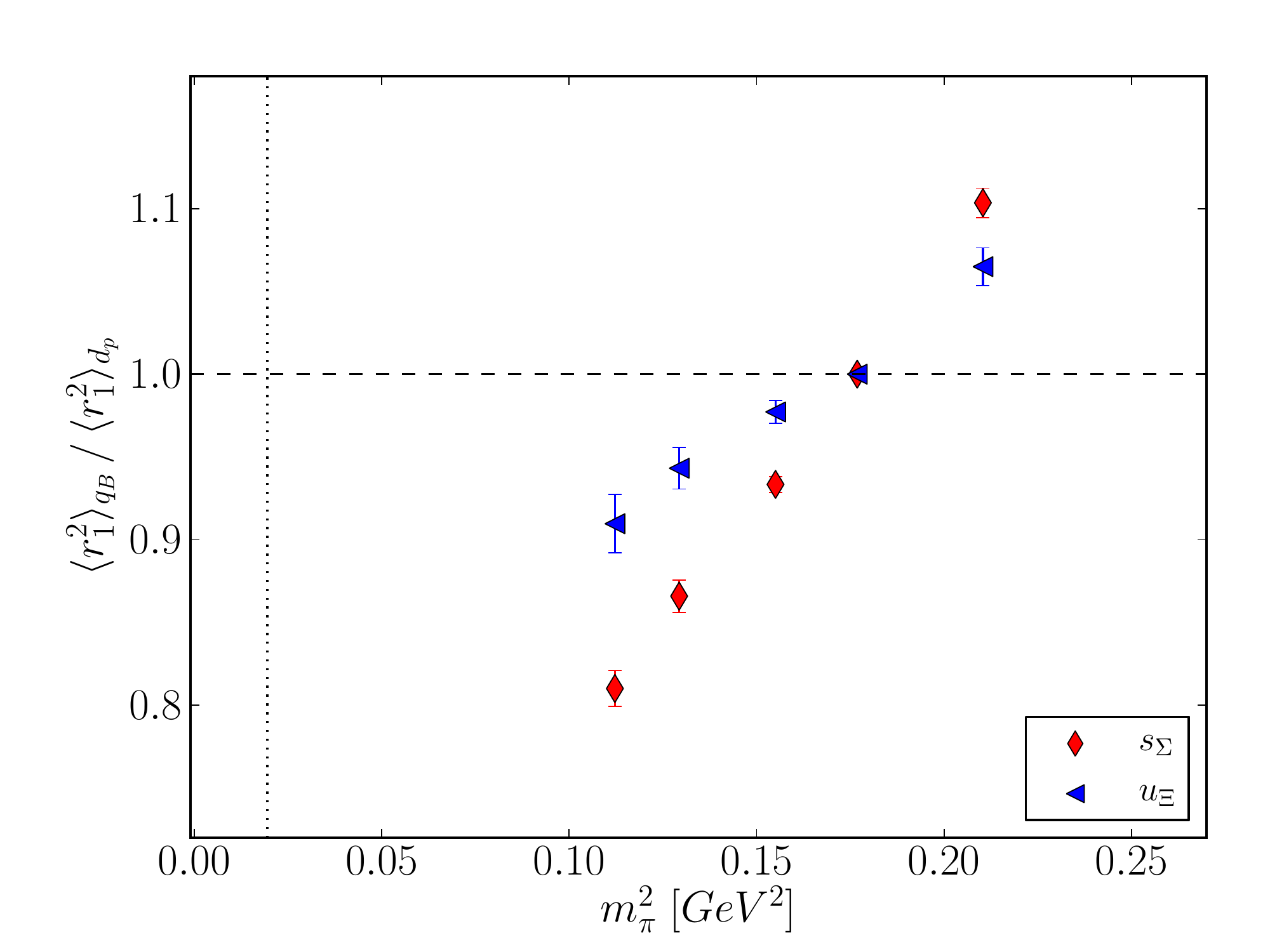}
          \caption{Results for the ratio of the singly-represented
            quark's contribution to the Dirac radius of the hyperons,
            $\langle r^2_1\rangle_{q_B}/\langle
            r^2_1\rangle_{d_p}$.}
\label{fig:r1S}
     \end{minipage}
 \end{figure}

In Fig.~\ref{fig:r1D} we see results for the ratio of the
doubly-represented quark's contribution to the Dirac radius of the
hyperons plotted as a function of $m_\pi^2$.
 Here we clearly see that the Dirac radii, $\langle r_1^2\rangle$, of
 both $u(d)$-quark in the $\Sigma^{+(-)}$ and the $s$-quark in the
 $\Xi^{0/-}$ become smaller than that of the $u$-quark in the proton
 as we move away from the $SU(3)_{\mathrm{flavour}}$ symmetric point
 by decreasing (increasing) the light (strange) quark mass.
This is particularly interesting in the case of $u_\Sigma$, since here
the only difference between $u_\Sigma$ and $u_p$ is the mass of the spectator
quark ($d(u)$ in the proton (neutron), $s$ in the $\Sigma$), so the fact that the
ratio is $<1$ is a purely environmental effect.

In Fig.~\ref{fig:r1S} we see a similar picture for the distribution of the
singly represented quarks with the $d(u)$-quark in the proton
(neutron) having a larger Dirac radius than the $u(d)$-quark in the
$\Xi^{0(-)}$ and the $s$-quark in the $\Sigma$ having the smallest
Dirac radius.

These effects are similar to those seen in an earlier quenched QCD
simulation \cite{Boinepalli:2006xd}, however here the effects are
slightly enhanced due to the fact that the strength of the meson loops
is suppressed in quenched QCD \cite{Leinweber:2002qb}.

Similar results are found for $\langle r_2^2\rangle$, albeit with
larger statistical errors due to the fact that the results need to be
extrapolated from $q^2\ne 0$.

\section{Hyperon Semi-Leptonic Form Factors}

The matrix element for $SU(3)$-octet baryon semileptonic decays, $B\to
b\ell\nu$, in Euclidean space is given by
\begin{equation}
  \label{eq:slme}
  \langle b(p',s')| V_\mu(x)+A_\mu(x)|B(p,s)\rangle = 
\overline{u}_b(p',s')({\cal O}_\mu^V(q) + {\cal O}_\mu^A(q))u_B(p,s)\ ,
\end{equation}
where the vector and axial-vector transitions are each governed
by three form factors, namely the vector $(f_1)$, weak magnetism
$(f_2)$, induced scalar $(f_3)$, axial-vector $(g_1)$, weak
electricity $(g_2)$ and induced pseudoscalar $(g_3)$
\begin{eqnarray}
  \label{eq:OVA}
{\cal O}_\mu^V (q) &=&
 f_1(q^2) \gamma_\mu +
 f_2(q^2) \sigma_{\mu \nu} \frac{q_\nu}{M_{b}+M_B} + 
 f_3(q^2) i\frac{q_\mu}{M_{b}+M_B}\ , \\
  {\cal O}_\mu^A (q) &=&
 g_1(q^2) \gamma_\mu \gamma_5 +
 g_2(q^2) \sigma_{\mu \nu} \frac{q_\nu}{M_{b}+M_B}\gamma_5 + 
 g_3(q^2) i\frac{q_\mu}{M_{b}+M_B}\gamma_5\ .
\end{eqnarray}
The vector and axial-vector currents in Eq.~(\ref{eq:slme}) are
defined as $V_\mu(x)=\bar{u}(x)\gamma_\mu d(x)$ and
$A_\mu(x)=\bar{u}(x)\gamma_\mu \gamma_5 d(x)$ for $\Delta S=0$ decays,
and $V_\mu(x)=\bar{u}(x)\gamma_\mu s(x)$ and
$A_\mu(x)=\bar{u}(x)\gamma_\mu \gamma_5 s(x)$ for $\Delta S=1$ decays.

For a lattice calculation of hyperon beta decays, it is useful to
define the scalar form factor
\begin{equation}
  \label{eq:f0}
  f_0(q^2)=f_1(q^2) + \frac{q^2}{M_B^2 + M_b^2}f_3(q^2)\ ,
\end{equation}
which can be obtained from the divergence of the vector current,
$\langle b(p',s')|\partial_\mu V_\mu|B(p,s)\rangle=(M_b -
M_B)f_0(q^2)\bar{u}(p',s')u(p,s)$, and the linear combination
\begin{equation}
  \label{eq:gtilde}
  \tilde{g}_1(q^2) = g_1(q^2) - \frac{M_B - M_b}{M_B + M_b}g_2(q^2)\ .
\end{equation}

The scalar form factor (\ref{eq:f0}) can be obtained on the lattice at
$q^2_{\mathrm{max}}=(M_B-M_b)^2$ with high precision from the ratio
\cite{Hashimoto:1999yp}
\begin{equation}
  R(t',t) =
  \frac{G^{Bb}_4(t',t;\vec{0},\vec{0})G^{bB}_4(t',t;\vec{0},\vec{0})}
  {G^{BB}_4(t',t;\vec{0},\vec{0})G^{bb}_4(t',t;\vec{0},\vec{0})}
  \xrightarrow[t,(t'-t) \to \infty]{} |f_0(q^2_{\mathrm{max}})|^2\ ,
  \label{eq:fq2max}
\end{equation}
where, e.g., $G^{Bb}_4(t',t;\vec{0},\vec{0})$, is the zero
three-momentum lattice three-point function of the fourth component
of the vector current, $V_4$, inserted at time $t$ between the source
baryon, $B$, located at time $t=0$ and the sink baryon, $b$, at time
$t'$.
We note that $R(t',t)=1$ in the $SU(3)_{\mathrm{flavour}}$ symmetric
limit, hence any deviations from unity are purely due to
$SU(3)_{\mathrm{flavour}}$ symmetry breaking effects.

\begin{figure}[t]
     \vspace*{-7mm}
     \hspace{-0.2mm}
   \begin{minipage}{0.47\textwidth}
     \hspace{-6mm}
          \includegraphics[clip=true,width=1.12\textwidth]{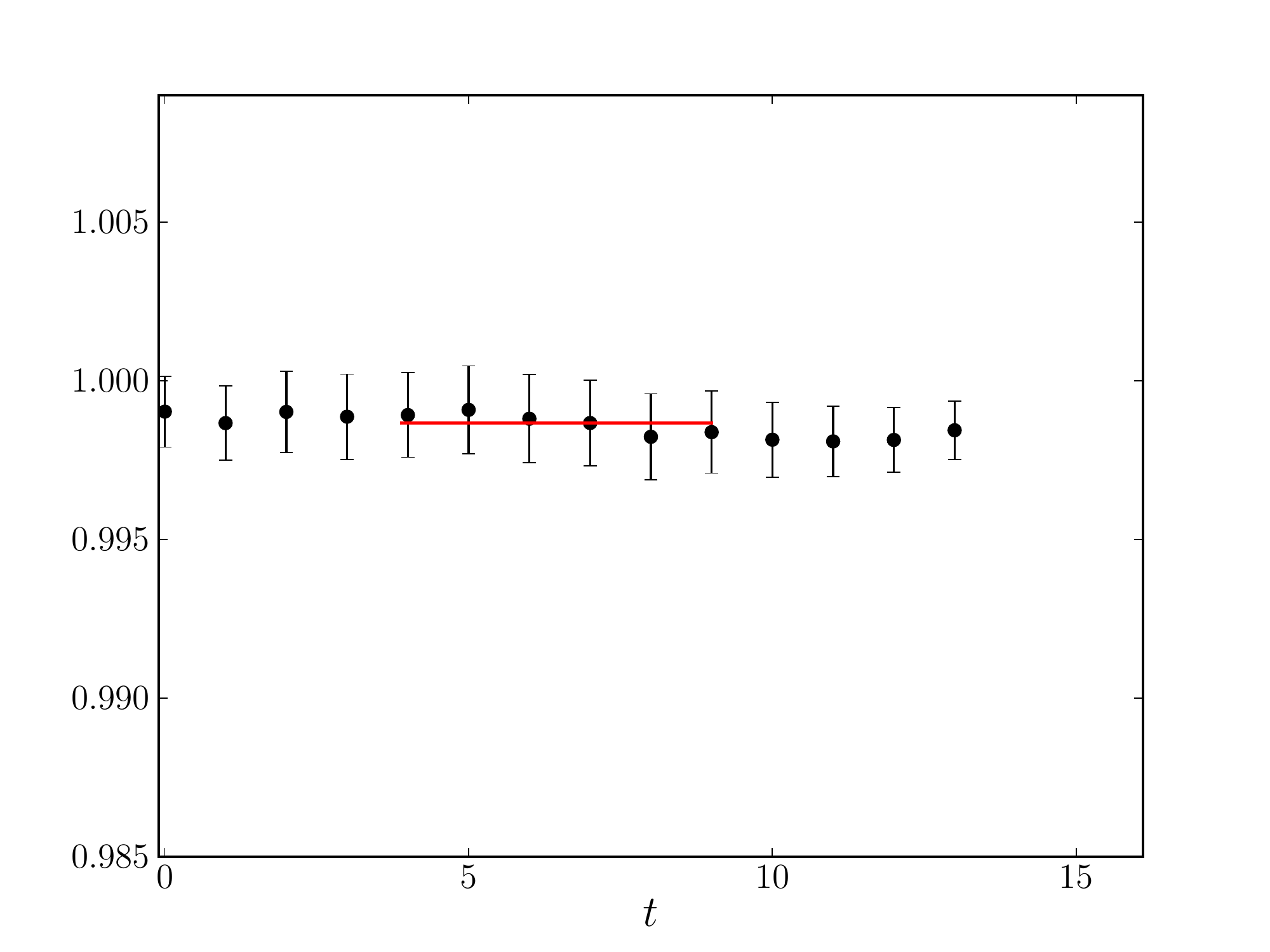}
          \caption{Ratio for $f_0(q^2_{\mathrm{max}})$, $R(t',t)$, as
            defined in Eq.~(\protect\ref{eq:fq2max}) for ensemble 4 for the
            $\Sigma^-\to n\ell\nu$ decay.}
\label{fig:s2n-q2max-rat}
     \end{minipage}
     \hspace{3mm}
    \begin{minipage}{0.47\textwidth}
     \hspace{-5mm}
          \includegraphics[clip=true,width=1.12\textwidth]{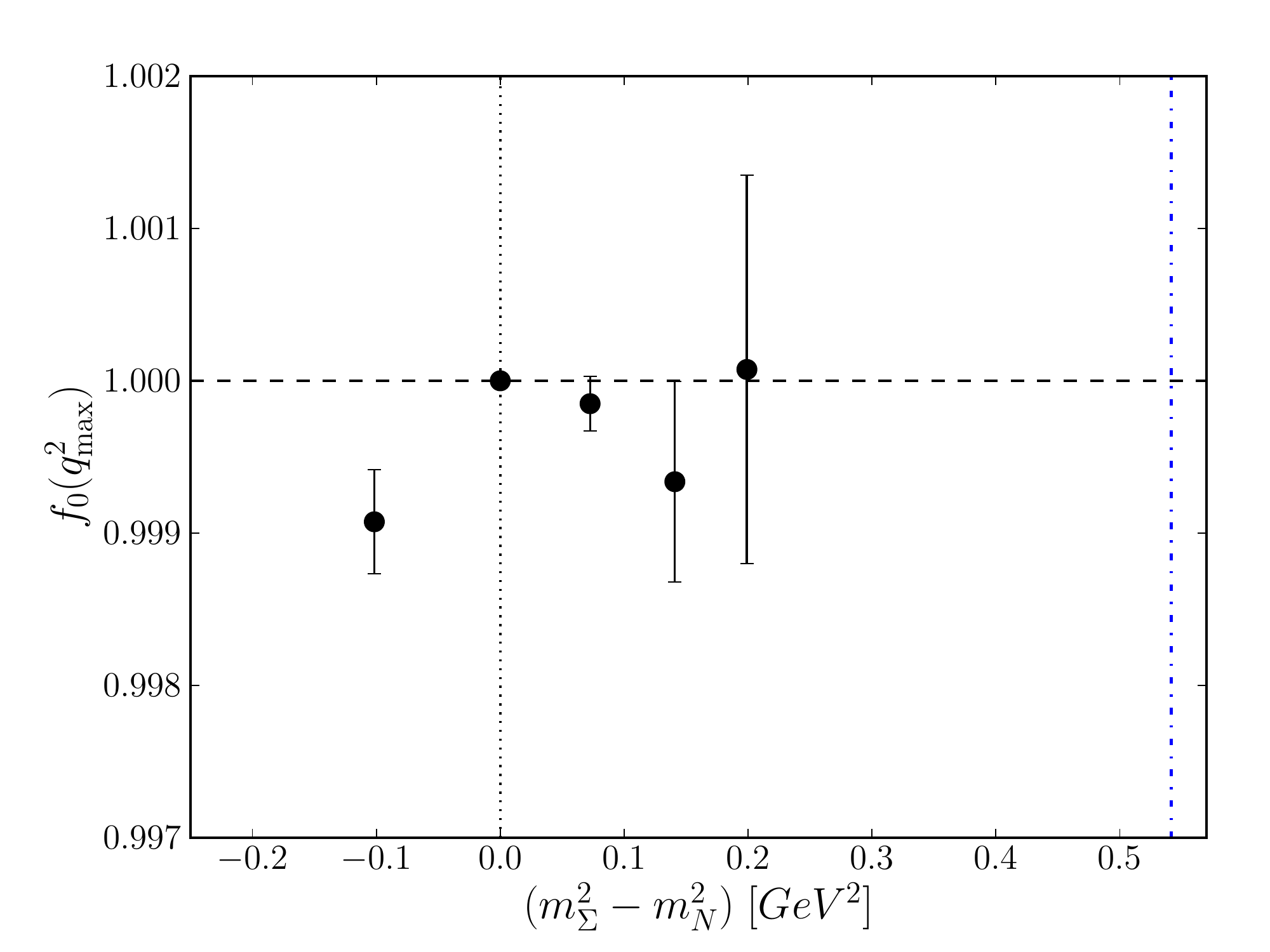}
          \caption{$f_0(q^2_{\mathrm{max}})$ for $\Sigma^-\to
            n\ell\nu$, as a function of the $\Sigma$-$n$ mass difference.}
\label{fig:s2n-q2max}
     \end{minipage}
 \end{figure}

In Fig.~\ref{fig:s2n-q2max-rat} we display our results for $R(t',t)$ for
ensemble 4 in Table~\ref{tab:results} for the $\Sigma^-\to n\ell\nu$
decay.
Here we see that it is possible to determine $f_0(q^2_{\mathrm{max}})$
with a high level of accuracy.
In order to quantify the size of the $SU(3)_{\mathrm{flavour}}$
symmetry breaking effects in the quantity, in Fig.~\ref{fig:s2n-q2max}
we show the results for $f_0(q^2_{\mathrm{max}})$ on each of our
ensembles.
The results are plotted as a function of $(m_\Sigma^2-m_N^2)$, hence
the $SU(3)_{\mathrm{flavour}}$ symmetric limit occurs at zero on the
$x$-axis and is indicated by the vertical dotted line, while the
physical mass splitting is indicated by the vertical dot-dashed line.
It is obvious from this figure that $f_0(q^2_{\mathrm{max}})$ is $<1$
away from the $SU(3)_{\mathrm{flavour}}$ symmetric limit and that the
deviation from unity increases as we move further away from the
$SU(3)_{\mathrm{flavour}}$ symmetric limit.
We find the same qualitative behaviour in our results for the
$\Xi^0\to\Sigma^+\ell\nu$ decay and these findings are in agreement
with earlier lattice results \cite{Guadagnoli:2006gj,Sasaki:2008ha}.

The next step is to determine the full $q^2$-dependence of $f_0(q^2)$
and interpolate the results to $q^2=0$ to obtain a value for
$f_1(0)=f_0(0)$.
The procedure for doing this has been described in detail in
\cite{Guadagnoli:2006gj}, and this is the next step in our work which
will be completed soon.

\begin{figure}[t]
     \vspace*{-7mm}
     \hspace{-0.2mm}
   \begin{minipage}{0.47\textwidth}
     \hspace{-6mm}
          \includegraphics[clip=true,width=1.12\textwidth]{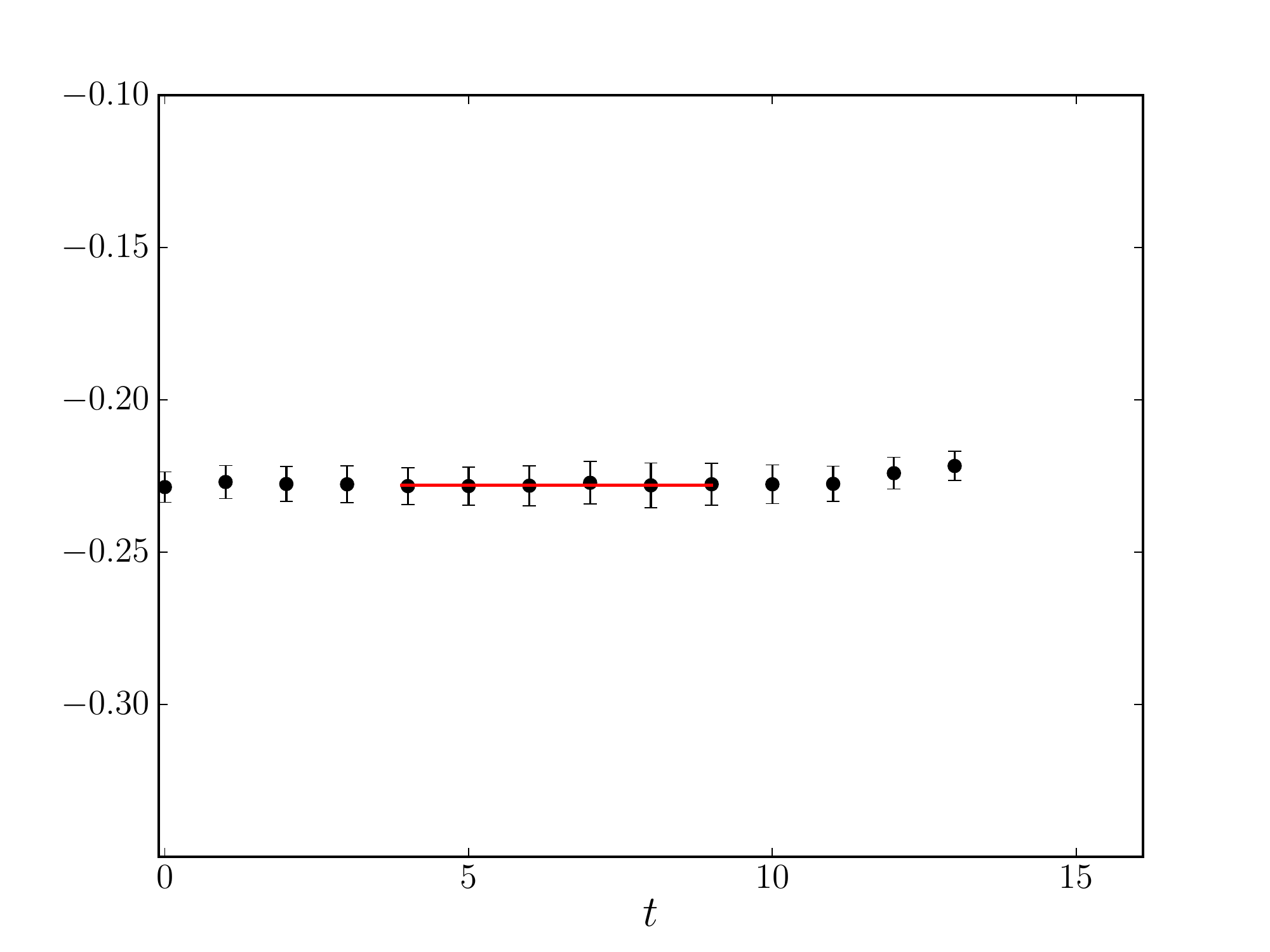}
          \caption{Ratio for
            $\tilde{g}_1(q^2_{\mathrm{max}})/f_0(q^2_{\mathrm{max}})$
            as defined in Eq.~(\protect\ref{eq:g-q2max-ratio}) for
            ensemble 1 for the $\Sigma^-\to n\ell\nu$ decay.}
\label{fig:s2n-slax-q2max-rat}
     \end{minipage}
     \hspace{3mm}
    \begin{minipage}{0.47\textwidth}
     \hspace{-5mm}
          \includegraphics[clip=true,width=1.12\textwidth]{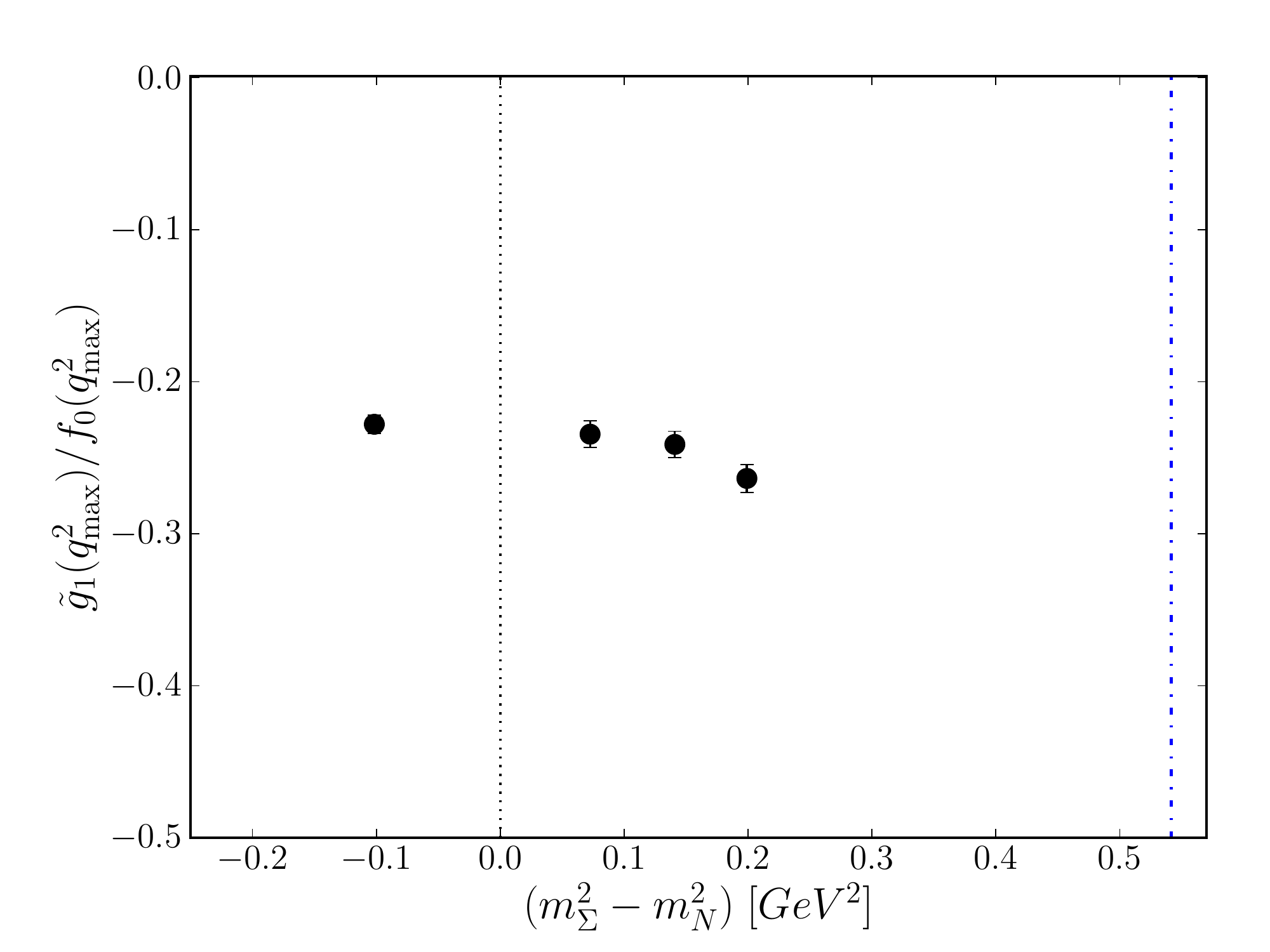}
          \caption{$\tilde{g}_1(q^2_{\mathrm{max}})/f_0(q^2_{\mathrm{max}})$
            for $\Sigma^-\to n\ell\nu$, as a function of the
            $\Sigma$-$n$ mass difference.}
\label{fig:s2n-slax-q2max}
     \end{minipage}
 \end{figure}

The other quantity that we need to calculate before we can determine
$|V_{us}|$ from Eq.~(\ref{eq:vus}) is $g_1(0)/f_1(0)$.
As pointed out in \cite{Guadagnoli:2006gj}, this can also be
determined from considering appropriate ratios of two- and three-point
functions.
At $q^2_{\mathrm{max}}$, it is possible to determine the ratio
$\tilde{g}_1(q^2_{\mathrm{max}})/f_0(q^2_{\mathrm{max}})$ from the
following ratio of three-point functions
\begin{equation}
  \label{eq:g-q2max-ratio}
  \tilde{R}(t',t) = 
  \frac{\mathrm{Im}\big(A_3^{Bb}(t',t;\vec{0},\vec{0})\big)}
  {\mathrm{Re}\big(V_4^{Bb}(t',t;\vec{0},\vec{0})\big)} 
  \xrightarrow[t,(t'-t) \to \infty]{}
  \frac{g_1(q^2_{\mathrm{max}}) +
    \frac{M_B-M_b}{M_B+M_b}g_2(q^2_{\mathrm{max}})} 
  {f_1(q^2_{\mathrm{max}})+\frac{M_B-M_b}{M_B+M_b}f_3(q^2_{\mathrm{max}})}
  \equiv
  \frac{\tilde{g}_1(q^2_{\mathrm{max}})}{f_0(q^2_{\mathrm{max}})}\ .
\end{equation}
We show in Fig.~\ref{fig:s2n-slax-q2max-rat} $\tilde{R}(t',t)$ from
ensemble 1 where we see again the excellent accuracy with which the
ratio can be determined.
The dependence of this ratio on the size of the
$SU(3)_{\mathrm{flavour}}$ symmetry breaking is seen in
Fig.~\ref{fig:s2n-slax-q2max}.

Once again, this is only the first step in calculating $g_1(0)/f_1(0)$
and so we must now subtract the second terms in both numerator and
denominator in Eq.~(\ref{eq:g-q2max-ratio}) in order to obtain
$g_1(q^2_{\mathrm{max}})/f_1(q^2_{\mathrm{max}})$.
We then need to map out the $q^2$-dependence of $g_1(q^2)/f_1(q^2)$
which will then enable us to interpolate to $q^2=0$.
This will be completed soon.

\section{Conclusions}

We have presented preliminary results from the QCDSF/UKQCD
collaboration for the electromagnetic semileptonic decay form factors
of the $SU(3)$ baryon octet.

Our results for the individual quark contributions to the Dirac radii
of the hyperons show that the $u(d)$-quark is more broadly distributed
in the proton (neutron) than in the $\Sigma^{+(-)}$, while the
$s$-quark in the $\Xi$ is the least broadly distributed of the
doubly-represented quarks.
Similarly for the singly-represented quarks, we find that the
$d(u)$-quark is more broadly distributed in the proton (neutron) than
in the $\Xi^{-(0)}$, while the $s$-quark in the $\Sigma$ is the least
broadly distributed.

For the hyperon semileptonic form factors, we have only performed the
first stage of the analysis by computing the appropriate form factors
at $q^2_{\mathrm{max}}$.
Our results are encouraging and show a similar quark mass behaviour
as earlier quenched \cite{Guadagnoli:2006gj} and $N_f=2$
\cite{Sasaki:2008ha} results.

\section*{Acknowledgements}

The numerical calculations have been performed on the apeNEXT at
NIC/DESY (Zeuthen, Germany), the IBM BlueGeneL at EPCC (Edinburgh,
UK), the BlueGeneL and P at NIC (J\"ulich, Germany), the SGI ICE 8200
at HLRN (Berlin-Hannover, Germany) and the JSCC (Moscow, Russia).
We thank all institutions.
We have made use of the Chroma software suite \cite{Edwards:2004sx},
employing the SSE optimised Dslash code \cite{SSE}, while our
Bluegene codes were optimised using Bagel \cite{Bagel}.
This work has been supported in part by the DFG (SFB/TR 55, Hadron
Physics from Lattice QCD) and the European Union under grants 238353
(ITN STRONGnet) and 227431 (HadronPhysics2).
JZ is supported through the UK's {\it STFC Advanced Fellowship
  Programme} under contract number ST/F009658/1.

\newpage
%
%

\end{document}